\documentclass[aps,twocolumn,preprintnumbers,amsmath,amssymb,prl]{revtex4-1}
\usepackage{graphicx}
\usepackage{dcolumn}
\usepackage{bm}
\usepackage{color}

\begin{document}




\title{Coexisting localized and itinerant gapless excitations  in a quantum spin liquid candidate 1T-TaS$_2$ }

\author{H.\,Murayama$^1$}
\author{Y.\,Sato$^1$}
\author{X.Z.\,Xing$^1$}
\author{T.\,Taniguchi$^1$}
\author{S.\,Kasahara$^1$}
\author{Y.\,Kasahara$^1$}
\author{M.\,Yoshida$^2$}
\author{Y.\,Iwasa$^{2,3}$}
\author{Y.\,Matsuda$^1$}

\affiliation{$^1$ Department of Physics, Kyoto University, Kyoto 606-8502 Japan}
\affiliation{$^2$ RIKEN Center for Emergent Matter Science (CEMS), Wako 351-0198, Japan} 
\affiliation{$^3$Quantum-Phase Electronics Center and Department of Applied Physics, the University of Tokyo, Tokyo 113-8656, Japan}




\begin{abstract}

To reveal the nature of elementary excitations in a quantum spin liquid (QSL), we measured  low temperature thermal conductivity  and specific heat of 1T-TaS$_2$, a QSL candidate material with frustrated triangular lattice of spin-1/2.  The nonzero  temperature linear specific heat coefficient $\gamma$  and the finite residual linear term of the thermal conductivity  in the zero temperature limit $\kappa_0/T=\kappa/T(T\rightarrow 0)$ are clearly resolved.  This demonstrates the presence of highly mobile gapless excitations, which is consistent with fractionalized spinon excitations that form a Fermi surface.  Remarkably, an external magnetic field strongly suppresses $\gamma$,  whereas it enhances $\kappa_0/T$.   This unusual contrasting behavior in the field dependence of  specific heat and thermal conductivity can be accounted for by  the presence of two types of gapless excitations with itinerant and localized characters, as recently predicted  theoretically (I. Kimchi {\it et al.}, arXiv:1803.00013 (2018)).  This unique feature of 1T-TaS$_2$ provides new insights into the influence of quenched disorder on the QSL.

\end{abstract}
\maketitle

Spin liquid is a state in which the constituent spins are highly correlated but continue to fluctuate strongly like in a liquid down to very low temperatures. In particular, quantum spin liquid (QSL) is a novel state of matter in which enhanced quantum fluctuations prevent the system from the long-range magnetic ordering  even at zero temperature \cite{spinon_Balents,Zhou2017}.  The ground states of QSLs are quantum-mechanically entangled and have characteristic features such as no simple symmetry breaking,  fractionalized excitations, and non-trivial topology.  The notion of QSLs is firmly established in one-dimensional (1D) spin systems as well as in their ladder cousins.  It is widely believed that in 2D and 3D systems, QSLs are usually realized in the presence of competing orders or geometrical frustration.     Despite tremendous efforts during the past several decades, however, the nature of QSLs in 2D and 3D remains mysterious.  One of the most important and  long-standing open problems,  which is a key  for understanding the elusive QSL states,  is how quantum fluctuations respond to quenched disorder from defects/impurities, including randomness of the magnetic exchange interaction.

Of specific interest in 2D frustrated spin systems has been the spin-1/2 triangular-lattice Heisenberg antiferromagnet, in which the very prototype of a QSL in resonating valence bond model with a quantum superposition of spin singlets,  has been proposed \cite{RVB, RVB2}.  Unfortunately, only a few candidates of the QSL state have been reported for 2D triangular lattice, including organic Mott insulators, $\kappa$-(BEDT-TTF)$_2$Cu$_2$(CN)$_3$ \cite{BEDT-TTF_NMR,BEDT-TTF_TC}, EtMe$_3$Sb[Pd(dmit)$_2$]$_2$ \cite{dmit_NMR1, dmit_NMR2} (hereafter abbreviated as DMIT) and  $\kappa$-H$_3$(Cat-EDT-TTF)$_2$ (H-Cat) \cite{Hcat,HCat2,HCat3} and inorganic YbMgGaO$_4$ \cite{Yb_C, Yb_crystal,Yb_spinon}. In particular, in the above organic systems, no magnetic ordering occurs at least down to 1/1000 of $J/k_B=$200-300\,K ($J$ is the exchange interaction between neighboring spins) \cite{BEDT-TTF_NMR, dmit_NMR1}.  The most remarkable and intriguing feature in the QSL state on triangular lattice is that  in DMIT and H-Cat the specific heat has a non-zero  linear temperature term $\gamma T$  and thermal conductivity has a finite residual  linear term $\kappa_0/T \neq0$.  This demonstrates the emergence of gapless spin excitations that behave like mobile charge carriers in a paramagnetic metal with a Fermi surface although the charge degrees of freedom are gapped \cite{dmit_k, dmit_C,HCat2,HCat3}.  The observed highly mobile  quasiparticles have  been discussed in terms of fermionic spinons, fractionalized particles that carry spin but no charge, and gapless  excitations have been attributed to a spinon Fermi surface \cite{Motrunich2005,Lee2005}.  However,  the triangular lattices consisting of  large molecular dimers with spin-1/2 in the above organic compounds are distorted from the perfect triangular lattice.  Moreover, it has been suggested that the features of the spin frustrations may be influenced by other factors, such as 
 freezing of the electric polarization degrees of freedom in the dimers \cite{Watanabe2014}.  On the other hand,   in YbMgGaO$_4$, $J/k_B$  is of order of 4\,K \cite{Yb_C, Yb_crystal,Yb_spinon}, making it difficult to extract intrinsic properties of the QSL state at low enough temperature, $T \ll J/k_B$.  Thus the nature of the gapless excitations in the QSLs on triangular lattice remains unclear and the situation calls for structurally perfect 2D triangular systems with large exchange interactions, which are free from the above issues.

\begin{figure}[t]
	\begin{center}
		\includegraphics[width=0.8\linewidth]{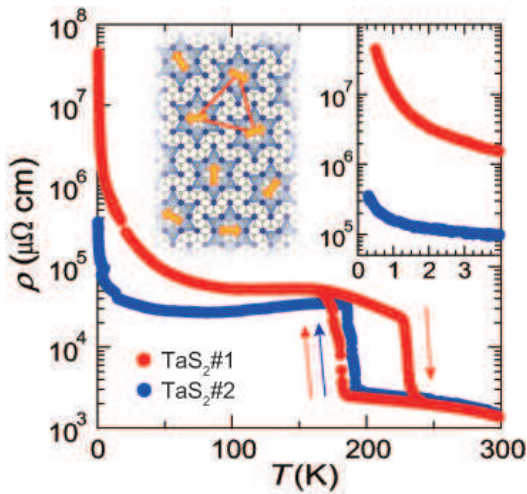}
		\caption{Temperature dependence of the resistivity of \#1 and \#2 TaS$_2$ single crystals. For \#1 crystal, the resistivity  measured both on cooling and on heating are shown. The right inset shows the low-temperature upturn in the resistivity. The left inset is the spin structure in the 2D planes of 1T-TaS$_2$ below C-CDW transition at 180\,K.  The unit cell is reconstructed into a rotated  triangular lattice characterized by $\sqrt{13} \times \sqrt{13}$ structure described as star-of-David clusters with 13 Ta atoms.   One  localized electron resides at the center of the star-of-David cluster and as a result, perfect 2D triangular lattice with $S=1/2$ is formed.
		}
		\label{fig:figure1}
	\end{center}
\end{figure}

Recently, transition metal dichalcogenide 1T-TaS$_2$ has aroused great interest as a candidate material that hosts a QSL ground state on the 2D perfect triangular lattice \cite{TaS2_Law}. This compound is a layered material that contains one Ta layer sandwiched by two S layers; these layers are weakly coupled by van der Waals interactions. The Ta atoms form a 2D triangular lattice. At high temperatures ($T>550$\,K), 1T-TaS$_2$ is metallic.  As the temperature is lowered,  it becomes an incommensurate CDW phase below 550\,K,  followed by a nearly commensurate CDW (NC-CDW) phase below 350\,K.   It undergoes a commensurate CDW (C-CDW) transition at 180\,K, below which the unit cell is reconstructed into a rotated  triangular lattice characterized by $\sqrt{13} \times \sqrt{13}$ structure described as star-of-David clusters with 13 Ta atoms \cite{TaS2_CDW, TaS2_C_chi}.   Strong electron correlation gives rise to a Mott insulating state, in which  one  localized electron resides at the center of the star-of-David cluster.  As a result, perfect 2D triangular lattice with $S=1/2$ is formed at low temperatures, as illustrated in the inset of Fig.\,1. 

Muon spin relaxation and nuclear quadrupole resonance (NQR) experiments have reported no long-range magnetic ordering down to 20\,mK, despite the large exchange coupling $J/k_B>1,000$\,K \cite{TaS2_muon_NQR}.   Moreover, gapless behavior of the spin dynamics has  also been suggested by these measurements.   To obtain information on the low-lying elementary excitations, the specific heat  and thermal conductivity  have been measured.   The specific heat contains contributions from  localized and itinerant excitations, while  the thermal conductivity is sensitive exclusively to itinerant excitations. Therefore the combination of both quantities provide pivotal information on the nature of the low energy quasiparticles.   The linear temperature term in the specific heat $\gamma T$  is clearly resolved, demonstrating the presence of gapless excitations \cite{TaS2_muon_C}, similar to DMIT and H-Cat \cite{dmit_C,HCat3}.  However, in stark contrast to these organic compounds, $\gamma(H)$ is strongly suppressed by external magnetic field.   The measurements of the  thermal conductivity report the absence of the residual linear term in zero field, $\kappa_0/T=\kappa/T (T\rightarrow 0)=0$, indicating the absence of itinerant magnetic excitations \cite{TaS2_k}.  Moreover, the thermal conductivity exhibits no  magnetic field dependence.  Based on the thermal conductivity, they concluded that there is no significant contribution of  itinerant quasiparticle excitations to the  thermal conductivity, which is again in stark contrast to the organic triangular compounds with finite $\kappa_0/T$, indicating no evidence of the spinon excitations.   These results suggest that the properties of the QSL state in 1T-TaS$_2$ may be markedly different from those of previously reported QSL candidate materials. 

Here we report  low temperature thermal conductivity and specific heat measurements on high quality single crystals of 1T-TaS$_2$.  As reported previously \cite{TaS2_muon_C},  non-zero  temperature linear specific heat coefficient $\gamma$, which is strongly suppressed by $H$,  is  observed.  Contrary to the previous report \cite{TaS2_k},  however, a sizable temperature linear term in the thermal conductivity, $\kappa_0/T \neq 0$ is clearly resolved, indicating that the ground state  is a thermal metal, although the charge degrees of freedom are gapped.  Moreover, in contrast to Ref.\,\cite{TaS2_k},  $\kappa/T$ is enhanced by $H$.   
Based on these results,  we argue that highly mobile gapless excitations coexist with localized gapless excitations arising from the local spin moments in the QSL state of 1T-TaS$_2$. 

High quality 1T-TaS$_2$ single crystals  were grown by the chemical vapor transport method. Figure\,1 and its inset depict the temperature dependence of the resistivity $\rho(T)$ for two crystals (\#1 and \#2) of 1T-TaS$_2$. For \#1 crystal, $\rho(T)$ measured both on cooling and warming is plotted. The hysteresis around 200\,K is due to the first order phase transition between NC-CDW and C-CDW.  Both crystals show the insulating behavior at low temperatures. While the values of $\rho(T)$ of two crystals at $\sim 170$\,K just below the C-CDW transition are similar, $\rho$ for \#1 crystal is two orders of magnitude larger than that for \#2 crystal at very low temperatures.

\begin{figure}[t]
	\begin{center}
		\includegraphics[width=0.8\linewidth]{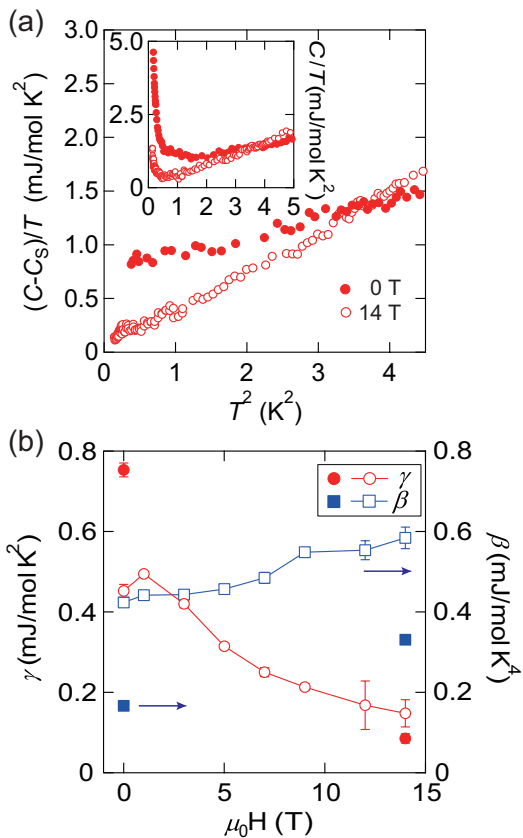}
		\caption{
			(a) The inset shows $C/T$ in zero field (filled red circles) and in magnetic field of $\mu_0H$=14\,T  for ${\bm H}\perp ab$ plane (open red circles) plotted  as a function of $T^2$. The main panel shows the  specific heat obtained after subtracting the Schottky  contribution, $(C-C_S)/T$, plotted  as a function of $T^2$ in zero field (filled red circles) and at  $\mu_0H$=14\,T (open red circles). (b) The $\gamma$-term (red filled circles) and $\beta$-term (blue filled squares) in the specific heat $C=\gamma T+\beta T^3+C_s$ in zero field  and at $\mu_0H=14$,T. The data represented by open symbols are taken from Ref.\,\cite{TaS2_muon_C}. 
		}
		\label{fig:figure2}
	\end{center}
\end{figure}

First we discuss the low-lying excitations in the absence of magnetic field.   The red filled circles in the  inset of Fig.\,2(a) shows $C/T$ plotted as a function of  $T^2$ of 1T-TaS$_2$ taken from the same batch of \#1 crystal.    As the temperature is lowered, $C/T$ decreases in proportion to $T^2$ in the high temperature regime.     Below $\sim$1.5\,K, $C/T$ shows a sharp upturn, which  is  attributed to the Schottky contribution $C_{\rm S}/T$.   Despite the presence of the Schottky anomaly, it is obvious that the extrapolation of $C/T$ above 1.5\,K to $T\rightarrow 0$ has nonzero intercept, indicating the presence of linear temperature term $\gamma T$, consistent with Ref.\,\cite{TaS2_muon_C}.  To analyze the low temperature specific heat, we subtract the Schottky term by assuming $C_{\rm S}/T\propto T^{-3}$.  As shown by the red filled circles in the main panel of Fig.\,2(a),  the low temperature specific heat  is well described as $(C-C_S)/T=\gamma+ \beta T^2$ with $\gamma\approx0.75$\,mJ/K$^2$mol and $\beta=0.17$\,mJ/K$^4$mol.  
Finite $\gamma$ demonstrates the presence of gapless excitations.  
 
We examine the thermal conductivity that provides the dynamical aspect of the  excitations.   The thermal conductivity is totally insensitive to localized entities that may cause the nuclear Schottky contribution and plague the heat capacity measurements at low temperatures.  The red and blue filled  circles in Figs.\,3(a) and 3(b) show $\kappa/T$ in zero field for \#1 and \#2 crystals, respectively,  plotted as a function of $T^2$.  The insets of Figs.\,3(a) and 3(b) depict the same data plotted as a function of $T$.     For either plot, non-zero intercepts of $\kappa/T$  extrapolated to $T\rightarrow 0$, i.e. $\kappa_0/T \neq 0$,  can be seen in both crystals.   Thermal conductivity in insulating magnets  can be written as a sum of the spin and phonon contributions, $\kappa=\kappa_{spin}+\kappa_{ph}$. The phonon conductivity in boundary-limit scattering regime at low temperature is expressed as $\kappa_{ph}=\frac{1}{3} p \langle v_s \rangle \ell_{ph}T^3$, where $p$ is the phonon specific heat coefficient, $\langle v_s \rangle$ is the mean acoustic phonon velocity, and $\ell_{ph}$ is the phonon mean free path.   For diffuse scattering limit, $\ell_{ph}$ becomes $T$-independent, resulting in $\kappa_{ph}\propto T^3$.  On the other hand, in case of specular reflection, $\ell_{ph}$  follows $T^{-1}$-dependence, leading to $\kappa_{ph}\propto T^2$.   In real systems, the phonon conductivity depends on $T$ as $\kappa_{ph}\propto T^{\alpha}$ with $\alpha$ of intermediate value between 2 and 3.   Therefore, the finite  $\kappa_0/T$ revealed by both plots of  $\kappa/T$ vs. $T^2$ and $\kappa/T$ vs. $T$, as shown in Figs.\,3(a) and 3(b) and their insets,  provides evidence of  finite temperature linear term in $\kappa_{spin}$, i.e. the presence of gapless {\it itinerant} spin excitations.  Such itinerant excitations in the QSLs have been attributed to emergent fractionalized quasiparticle ``spinon", which carries spin but no charge.  Moreover,  the gapless excitations represented by finite $\gamma$ and $\kappa_0/T$  are consistent with a spinon Fermi surface \cite{Motrunich2005,Lee2005,He_spinon}, ruling out  a Dirac spinon with nodes.

\begin{figure}[t]
	\begin{center}
		\includegraphics[width=1.0\linewidth]{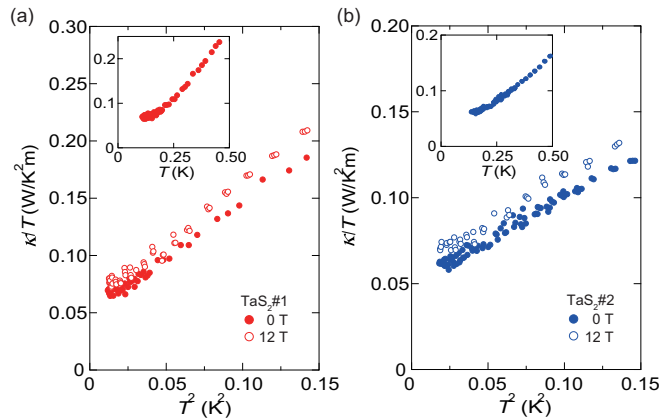}
		\caption{(a) Low temperature plot of $\kappa/T$  as a function of  $T^2$ of \#1 TaS$_2$ single crystal in zero magnetic field (filled circles) and at $\mu_0H=12$\,T for {\boldmath $H$} $\perp ab$ plane (open circles).   The inset shows $\kappa/T$ in zero field  plotted as a function of $T$.  (b) The same plots for \#2  crystal.}
		\label{fig:figure3}
	\end{center}
\end{figure}

While the resistivity of \#1 crystal is two orders of magnitude larger than that of \#2 at low temperatures, residual thermal conductivity of \#1 is very close to that of \#2.  This indicates that the mean free path of the itinerant spin excitations is not directly related to the electron hopping channel responsible for the electrical resistivity.  The present results are in contrast to  the previous measurements that report the absence of $\kappa_0/T$ \cite{TaS2_k}.  This discrepancy may be because the defects/impurities reduce $\kappa_0/T$ to a level beyond the resolution of the experiment.

Next we discuss the influence of the magnetic field on the gapless excitations. As shown in Fig.\,2(a) and its inset, the magnetic field of $\mu_0H=14$\,T applied perpendicular to the 2D plane dramatically changes the specific heat, which is consistent with the data reported in Ref.\,\cite{TaS2_muon_C}.  The low-temperature Schottky contribution is suppressed by magnetic field. After the subtraction of Schottky term, specific heat is represented as  $\frac{C-C_S}{T}(H)=\gamma(H) +\beta(H) T^2$.   The most salient features are that the magnetic field strongly suppresses $\gamma(H)$, while it largely enhances $\beta(H)$.  This indicates that $\beta$-term in zero field contains not only phonon contribution but also spin contribution.  It should be stressed that this highly unusual field dependence of the specific heat has never been observed in the other QSL candidate materials.  In Fig.\,2(b),  the $H$-dependences of $\gamma(H)$- and $\beta(H)$-terms of our results are shown by filled symbols. For comparison, $\gamma(H)$ and $\beta(H)$  reported in Ref.\,\cite{TaS2_muon_C} are  plotted by open symbol.  In zero field, the magnitude of $\gamma$- and $\beta$-terms  in our crystal are $\sim 1.7$ times larger and $\sim 2.5$ times smaller  than the previously reported values \cite{TaS2_muon_C}.   The present crystal exhibits larger suppression of  $\gamma$-term and larger enhancement of  $\beta$-term by magnetic field. In particular, $\beta(14\,{\rm T})/\beta(0)\approx 2$ in the present crystal is $\sim$1.5 time larger than that reported in Ref.\,\cite{TaS2_muon_C}.  We will discuss this difference later.   

The open circles in Figs.\,3 (a) and 3(b) show $T$-dependence of $\kappa/T$ of \#1 and \#2 crystals, respectively,  in the magnetic field of $\mu_0H=12$\,T applied perpendicular to the 2D plane.  In contrast to the specific heat,  the thermal conductivity is enhanced by the magnetic field.  As  the phonon mean free path is limited by the crystal size in this temperature regime, $\kappa_{ph}$ is not enhanced by magnetic field.  Therefore the enhancement of $\kappa/T$ at finite temperature is attributed to a pure spin sector  contributions.

\begin{figure}[t]
	\begin{center}
		\includegraphics[width=0.8\linewidth]{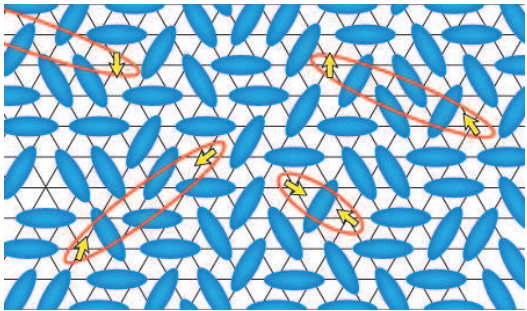}
		\caption{
			A schematic illustration of spin-1/2 defects in a QSL state.  Blue ellipsoids represent  resonating spin singlets, which give rise to itinerant excitations. Yellow arrows represent randomly distributed orphan spins induced by the quenched disorder from defects/impurities, including randomness in the exchange couplings. These orphan spins form localized random singlets (ellipses with red line), which give rise to  gapless spin excitations. 
		}
		\label{fig:figure4}
	\end{center}
\end{figure}

The observed contrasting behavior in  the field dependence of specific heat and thermal conductivity is highly unusual.  It should be stressed that this provides a key for understanding the nature of the gapless excitations in the QSL state of 1T-TaS$_2$.  Recently, it has been proposed that the large suppression of the $\gamma$-term in the specific heat by magnetic field indicates the presence of gapless excitations by local magnetic moments \cite{VBS_defect, QSL_defect_C}.    In this scenario, resonating spin singlets give rise to itinerant excitations in the QSL state, which can be gapless.  The weak randomness in $J$ induced by defects/impurities leads to the emergence of localized orphan spins that remain out of resonating singlets (see Fig.\,4).    These randomly distributed orphan spins form the localized random singlets, whose binding energy decrease with distance.  As a result, excitation energy  increases continuous from zero, i.e. gapless.   The presence of randomly distributed orphan spins is consistent with highly inhomogeneous magnetic state reported  in NQR \cite{TaS2_muon_NQR}.  The magnetic field breaks these localized spin singlets, leading to the suppression of  $\gamma$-term in the specific heat.  In 1T-TaS$_2$, deficiency of S can induce the weak randomness in $J$.   It should be stressed that the observed field induced enhancement of thermal conductivity is consistent with this scenario because the itinerant quasiparticles are less scattered by the orphan spins polarized by magnetic field. Moreover,  large differences of the magnitude and the field dependence of $\gamma$- and $\beta$-terms between the present crystal and the crystal of Ref.\,\cite{TaS2_muon_C} appears to be in line with the theory of localized random singlets, because both crystals should have different level of defects/impurities. To clarify this point, more systematic specific heat measurements are required.

The strong field dependence of  $\gamma$- and $\beta$-terms has not been observed in  organic triangular-lattice spin systems.  This may be owing to several intrinsic  differences between organic compounds and the present 1T-TaS$_2$.  First, the closeness to Mott transition between two systems is different due to the different charge gaps.  Then the influence of the  randomness on the quantum fluctuations may be  less pronounced in organic compounds.  Second,  these organic compounds may be cleaner than 1T-TaS$_2$, as  charge transfer salts  usually have less impurities and lattice defects.   Third, the unpaired electron with spin-1/2 in organic compounds occupies  in a large molecular dimer and hence may be less influenced by local defects.   Further studies are required to clarify these issues. 

We point out that the present results do not support the phase separation scenario, in which spatially separated  glassy and QSL phases coexists.  In this scenario the strong reduction of $\gamma$-term occurs in the former phase and the thermal conductivity is governed by the latter phase. However,  it is  unlikely that $\kappa_0/T$  is enhanced for nearly 10-15\% by magnetic field of $\mu_0H=12$\,T, which is less than 1\% of $J$. Moreover, such a scenario is  inconsistent to  the  NQR line shape that shows little broadening  at very low temperatures \cite{TaS2_muon_NQR}.     

Here we estimate the mean free path of the gapless excitations. As shown in Fig.\,2(b), $\gamma$  shows the tendency of saturation at 14\,T, suggesting that the orphan spins are nearly fully polarized.  This indicates that the $\gamma$ at 14\,T is dominated by the itinerant gapless excitations.    Assuming the kinetic approximation, the thermal conductivity is written as $\kappa_{0}/T=\frac{1}{3}\gamma_s v_s\ell_s$, where $\gamma_s$ is the specific heat coefficient per spin, $\gamma_s=13\gamma$,  $v_{s}$ is the velocity and $\ell_{s}$ is the mean free path of the quasiparticles responsible for the gapless excitations.  We  estimate $\ell_s$ simply assuming that the linear term in the thermal conductivity arises from the gapless fermionic excitations, in analogue to excitations near the Fermi surface in metals.  Using $J\sim$0.13 eV reported from the magnetic susceptibility measurements \cite{TaS2_muon_NQR} and  $v_{s}\sim \frac{Ja}{\hbar}$, where $a=$1\,nm is the inter-spin distance, we obtain  $\ell_s\sim$20.4\,nm.  This mean free path, which is much longer  than inter-spin distance, is consistent with the itinerant nature of the gapless excitations.  

 In summary, we have measured the thermal conductivity and specific heat at low temperatures in the QSL candidate material 1T-TaS$_2$ with perfect  triangular lattice of spin 1/2.  We find a finite $\gamma$-term in the specific heat and a finite  $\kappa_0/T$ in the thermal conductivity, demonstrating the presence of gapless fermionic excitations.  While an external magnetic field strongly suppresses $\gamma(H)$,  it enhances $\kappa_0/T(H)$.   These highly unusual contrasting response to magnetic field can be accounted for by the presence of two types of gapless excitations with itinerant and localized characters. The itinerant gapless excitations in the QSLs are consistent with emergent spinons that form a Fermi surface.   Accordingly, the itinerant spinons coexist with randomly distributed orphan spins forming the localized random singlets.  This  provides  new important  insights in understanding the effects of quenched disorder on the quantum fluctuations in the QSL.
 
\begin{acknowledgments}
We are very grateful to I. Kimchi, K. T. Law, P.A. Lee, E.-G. Moon,  K. Totsuka and M. Udagawa for many useful comments and suggestions. This work was supported by Grants-in-Aid for Scientific Research (KAKENHI) (Nos. 25220710, 25000003, 15H02014, 15H02106, and 15H05457) and on Innovative Areas `Topological Material Science' (No. JP15H05852) and `3D Active-Site Science' (No. 26105004) from Japan Society for the Promotion of Science (JPSJ). 
\end{acknowledgments}


\newpage
\begin{thebibliography}{5}
\bibitem{spinon_Balents}L. Savary and L. Balents, Quantum Spin Liquids. Rep. Prog. Phys. {\bf 80}, 016502 (2017). 
\bibitem{Zhou2017}Y. Zhou, K. Kanoda, and T-K Ng, Quantum spin liquid states. Rev. Mod. Phys. {\bf 89}, 025003 (2017).
\bibitem{RVB} P. W. Anderson, Resonating valence bonds: A new kind of insulator? Mat. Res. Bull. {\bf8,} 153 (1973).
\bibitem{RVB2}P. Fazekas and P. W. Anderson, On the ground state properties of the anisotropic triangular antiferromagnet. Philos. Mag. {\bf30,} 423 (1974).
\bibitem{BEDT-TTF_NMR}Y. Shimizu, K. Miyagawa, K. Kanoda, M. Maesato, G. Saito, Spin liquid state in an organic Mott insulator with a triangular lattice. Phys. Rev. Lett. {\bf 91}, 107001  (2003).
\bibitem{BEDT-TTF_TC}M. Yamashita, N. Nakata, Y. Kasahara, T. Sasaki, N. Yoneyama, N. Kobayashi, S. Fujimoto, T. Shibauchi, and Y. Matsuda, Thermal-transport measurements in a quantum spin-liquid state of the frustrated triangular magnet $\kappa$-(BEDT-TTF)$_2$Cu$_2$(CN)$_3$, Nat. Phys. {\bf 5}, 44-47 (2009).
\bibitem{dmit_NMR1}T. Itou, A. Oyamada, S. Maegawa, M. Tamura, and R. Kato, Quantum spin liquid in the spin-1/2 triangular antiferromagnet EtMe$_3$Sb[Pd(dmit)$_2$]$_2$. Phys. Rev. B {\bf77,} 104413 (2008).
\bibitem{dmit_NMR2}T. Itou, A. Oyamada, S. Maegawa, R. Kato, Instability of a quantum spin liquid in an organic triangular-lattice antiferromagnet. Nat. Phys. {\bf 6}, 673 (2010).
\bibitem{Hcat}T. Isono, H. Kamo, A. Ueda, K. Takahashi, M. Kimata, H. Tajima, S. Tsuchiya, T. Terashima, S. Uji, and H. Mori, Gappless Quantum Spin Liquid in an Organic Spin-1/2 Triangular-Lattice $\kappa$-H$_3$(Cat-EDT-TTF)$_2$. Phys. Rev. Lett. {\bf 112}, 177201 (2014). 
\bibitem{HCat2}M. Shimozawa, K. Hashimoto, A. Ueda, Y. Suzuki, K. Sugii, S. Yamada, Y. Imai, R. Kobayashi, K. Itoh, S. Iguchi, M. Naka, S. Ishihara, H. Mori, T. Sasaki and  M. Yamashita, Quantum-disordered state of magnetic and electric dipoles in an organic Mott system.  Nat. Commun. {\bf 8}, 1821 (2017).
\bibitem{HCat3}S. Yamashita, Y. Nakazawa, A. Ueda, and H. Mori, Thermodynamics of the quantum spin liquid state of the single-component dimer Mott system $\kappa$-H$_3$(Cat-EDT-TTF)$_2$. Phys. Rev. B {\bf 95}, 184425 (2017).
\bibitem{Yb_C}Y. Li, H. Liao, Z. Zhang, S. Li, F. Jin, L. Ling, L. Zhang, Y. Zou, L. Pi, Z. Yang, J. Wang, Z. Wu, Q. Zhang, Gapless quantum spin liquid ground state in the two dimensional spin-1/2 triangular antiferromagnet YbMgGaO$_4$. Sci. Rep. {\bf5}, 16419 (2015).
\bibitem{Yb_crystal}Y. Li, G. Chen, W. Tong, L. Pi, J. Liu, Z. Yang, X. Wang, Q. Zhang, Rare-Earth Triangular Lattice Spin Liquid: A Single-Crystal Study of YbMgGaO$_4$. Phys. Rev. Lett. {\bf115}, 167203 (2015).
\bibitem{Yb_spinon}Y. Shen, Y. Li, H. Wo, Y. Li, S. Shen, B. Pan, Q. Wang, H. C. Walker, P. Steffens, M. Boehm, Y. Hao, D. L. Quintero-castro, L. W. Harriger, M. D. Frontzek, L. Hao, S. Meng, Q. Zhang, G. Chen, J. Zhao, Evidence for a spinon Fermi surface in a triangularlattice quantum-spin-liquid candidate. Nature {\bf 540}, 559 (2016).
\bibitem{dmit_k}M. Yamashita, N. Nakata, Y. Senshu, M. Nagata, H. M. Yamamoto, R. Kato, T. Shibauchi, Y. Matsuda, Highly Mobile Gapless Excitations in a Two-Dimensional Candidate Quantum Spin Liquid. Science {\bf328}, 1246 (2010).
\bibitem{dmit_C}S. Yamashita, T. Yamamoto, Y. Nakazawa, M. Tamura, R. Kato, Gapless spin liquid of an organic triangular compound evidenced by thermodynamic measurements. Nat. Commun. {\bf2}, 275 (2011).
\bibitem{Motrunich2005}O. I. Motrunich, Variational study of triangular lattice spin-1/2 model with ring exchanges and spin liquid state in $\kappa$-(ET)$_2$Cu$_2$(CN)$_3$. Phys. Rev. B {\bf 72}, 045105 (2005).
\bibitem{Lee2005}S.-S. Lee, P. A. Lee, U(1) Gauge Theory of the Hubbard Model: Spin Liquid States and Possible Application to $\kappa$-(BEDT-TTF)$_2$Cu$_2$(CN)$_3$. Phys. Rev. Lett. {\bf 95}, 036403 (2005).
\bibitem{Watanabe2014}K. Watanabe, H. Kawamura, H. Nakano, and T. Sakai, Quantum Spin-Liquid Behavior in the Spin-1/2 Random Heisenberg Antiferromagnet on the Triangular Lattice. J. Phys. Soc. Jph. {\bf 83}, 034714 (2014).
\bibitem{TaS2_Law}K. T. Law, P. A. Lee, 1T-TaS$_2$ as a quantum spin liquid. Proc. Natl., Acad. Sci. USA {\bf114}, 6996 (2017).
\bibitem{TaS2_CDW}K. Rossnagel, On the origin of charge-density waves in select layered transition-metal dichalcogenides. J. Phys.: Condens. Matter {\bf23,} 213001 (2011).
\bibitem{TaS2_C_chi}M. Kratochvilova, A. D. Hillier, A. R. Wildes, L. Wang, S. Cheong, J. Park, The low-temperature highly correlated quantum phase in the charge-density-wave 1T-TaS$_2$ compound. npj Quantum Materials {\bf2}, 42 (2017).
\bibitem{TaS2_muon_NQR}Martin Klanj\v{s}ek, Andrej Zorko, Rok \v{Z}itko, Jernej Mravlje, Zvonko Jagli\v{c}i\'{c}, Pabitra Kumar Biswas, Peter Prelov\v{s}ek, Dragan Mihailovic, Denis Ar\v{c}on, A high-temperature quantum spin liquid with polaron spins. Nat. Phys. {\bf13,} 1130 (2017).
\bibitem{TaS2_muon_C}A. Ribak, I. Silber, C. Baines, K. Chashka, Z. Salman, Y. Dagan, A. Kanigel, Gapless excitations in the ground state of 1T-TaS$_2$. Phys. Rev. B {\bf96,} 195131 (2017).
\bibitem{TaS2_k}Y. J. Yu, Y. Xu, L. P. He, M. Kratochvilova, Y. Y. Huang, J. M. Ni, Lihai Wang, Sang-Wook Cheong, Je-Geun Park, S. Y. Li, Heat transport study of the spin liquid candidate 1T-TaS$_2$. Phys. Rev. B {\bf96,} 081111 (2017).
\bibitem{He_spinon}W.-Y. He, X. Y. Xu, G. Chen, K. T. Law, and P. A. Lee, Spinon Fermi surface in a cluster Mott insulator model on a triangular lattice and possible application to 1T-TaS$_2$. arXiv: 1803.00999 (2018).
\bibitem{VBS_defect}I. Kimchi, A. Nahum, T. Senthil, Valence Bonds in Random Quantum Magnets: Theory and Application to YbMgGaO$_4$. arXiv:1710.06860 (2017).
\bibitem{QSL_defect_C}I. Kimchi, J. P. Sheckelton, T. M. McQueen, P. A. Lee, Heat capacity from local moments in frustrated disordered quantum spin systems: scaling and data collapse. arXiv:1803.00013 (2018).



\end{thebibliography}
\end{document}